# Synthesis of hydrogen- and methyl-capped long-chain polyynes by intense ultrashort laser pulse irradiation of toluene

Ali Ramadhan[1]*[†], Michal Wesolowski[2]*[†], Tomonari Wakabayashi[3], Haruo Shiromaru[4], Tatsuya Fujino[4], Takeshi Kodama[4], Walter Duley[1], Joseph Sanderson[1]

[1] Department of Physics and Astronomy, University of Waterloo, 200 University Avenue West, Waterloo, Ontario, Canada N2L 3G1
[2] Department of Medical Imaging, University of Saskatchewan, 103 Hospital Drive, Saskatoon, Saskatchewan, Canada S7N 0W8
[3] Department of Chemistry, Kindai University, Kowakae 3-4-1, Higashi-Osaka 577-8502, Japan
[4] Department of Chemistry, Tokyo Metropolitan University, Minami-Osawa 1-1, Tokyo 192-0397, Japan

**Abstract**

Hydrogen- and methyl-capped polyynes have been synthesized by irradiating pure liquid toluene with 35 femtosecond, 300 μJ laser pulses having a central wavelength of 800 nm, generated by a regeneratively amplified Ti:sapphire tabletop laser at a repetition rate of 1 kHz. Raman spectroscopy was used to confirm the presence of polyynes in the irradiated samples while high-performance liquid chromatography was used to separate hydrogen-capped polyynes up to $C_{18}H_2$ and methyl-capped polyynes up to $HC_{14}CH_3$. These represent the first such methyl-capped polyynes and the longest hydrogen capped chains synthesized to date by the ultrafast laser based method. Furthermore our results show that choice of the starting solvent molecule directly influences the end caps of the polyynes which can be produced.

## 1. Introduction

Polyynes are molecules consisting entirely of sp-hybridized carbon bonds with chemical formula $(-C\equiv C-)_n$ where n > 1. They may be capped on either end, by hydrogen or larger end groups, which serve to stabilize longer chains [1], modify electronic and optical properties [2], or grant exotic properties [3]. Polyynes may possibly be used as nano-conductors as they can form novel coaxial nano-wires by insertion into single wall carbon nanotubes [4] and have also been observed in nanostructured thin films [5]. Carbyne, a theoretical allotrope of carbon consisting of infinite polyyne chains, has been predicted to be stable at room temperature with a specific strength twice that of graphene. With appropriate $CH_2$ end caps the carbon chain may be endowed with nonzero tensile stiffness and with a further 90° twist it may also switch into a magnetic semiconductor state [3].

* Corresponding authors. Emails: ali.ramadhan@uwaterloo.ca (Ali Ramadhan) and mike.wesolowski@usask.ca (Michal Wesolowski).
[†] These authors contributed equally to this work.

While the traditional method used to synthesize polyynes, oxidative coupling, has recently been used to generate long chain polyynes (n = 22) with stabilizing endcaps [1], the synthesis of carbyne has yet to be realized. Of particular note is that organic synthesis has also been successful for producing macroscopic quantities of cyanopolyynes such as $HC_5N$ [6]. Oxidative coupling itself remains a challenging, multistep procedure that can be extremely hazardous [7]. Recent progress on synthesizing carbyne has involved the growth of linear carbon chains in carbon nanotubes. Shi et al. [8] were able to synthesize polyynes thousands of carbon atoms long by using high vacuum alcohol chemical vapor deposition to synthesize double-walled carbon nanotubes then using high temperature high vacuum annealing to grow the linear carbon chains inside them. Kang et al. [2] synthesized linear carbon chains in multi-walled carbon nanotubes using atmospheric arc discharge in the presence of boron. However, the growth mechanisms employed are not fully understood yet and so it is unclear as to how the end caps may be modified. Recently, two methods involving pulsed laser irradiation have been successfully used to produce polyynes. The first involves the nanosecond laser irradiation of suspended carbon particles (e.g. graphite, fullerenes, and nano-diamond) in organic solvents [9] producing polyynes up to $C_{30}H_2$ in certain experiments [10]. In the second and more facile method, femtosecond laser radiation directly initiates synthesis through irradiation of an organic liquid without the need for the introduction of any carbon particles [11]. Sato et al. [12] have used this method to irradiate hexane, producing polyynes up to $C_{12}H_2$, and Wesolowski et al. [13] irradiated benzene, producing both amorphous carbon nanoparticles and polyynes up to $C_{14}H_2$; however, all synthesized polyynes were hydrogen-capped.

As the number of carbon atoms and the end caps can modify the electronic and optical properties of polyynes [2,3], the ability to control the chemical structure of the end caps is desirable. To date, this has not been demonstrated using laser based synthesis methods. To determine if the chemical structure of polyyne end caps can be influenced in laser based synthesis experiments and thereby expand on the potential utility of these polyyne synthesis techniques we have irradiated toluene (methylbenzene) with the aim of creating methyl-capped polyynes.

## 2. Experimental

High-performance liquid chromatography (HPLC) grade toluene (Sigma Aldrich CHROMASOLV®, ≥99.9% purity) in small glass vials (4.7 mL, base of 15 x 15 mm, height of 45 mm) was irradiated using 35 fs, 300 µJ laser pulses from a regeneratively amplified Ti:Sapphire tabletop laser with a 1 kHz repetition rate and central wavelength of 800 nm. Typically 1-1.5 mL of toluene was used. A schematic of this setup is shown in Fig. 1(a).

The laser beam was focused at the meniscus of the liquid using a mirror and a 50 mm focal length biconvex lens as shown in Fig. 1(a). Consequently, a very intense ($> 10^{13}$ W/cm$^2$), highly luminous, dissociation and ionization region formed intersecting the air-liquid boundary, visible in Figs. 1(b1) and 1(b2). Within a half hour after the start of the irradiation, the toluene formed a slight yellow hue, which continually darkened as the irradiation progressed. By the ninety-minute mark, a black precipitate began to form on the liquid meniscus starting at the walls and after three hours of irradiation covered most of the liquid surface. Physically disturbing the vial caused the precipitate to break up into macroscopic carbon particles that were free to move in the liquid due to the turbulence and convection caused by the energy released in the laser focus. The irradiation time was varied between 1 and 4 hours in 1 hour steps, however, sufficient polyyne concentrations were found only in the 3 and 4-hour samples. Photographs of the irradiated samples are shown in Fig. 1(c).

Irradiated samples were then characterized by Raman spectroscopy and high-performance liquid chromatography (HPLC). Raman spectroscopic measurements were carried out on a Renishaw micro-Raman Spectrometer with an objective magnification of 50 and an excitation wavelength of 488 nm (0.5 mW). For HPLC, a Shimadzu LC-10A liquid chromatographic system equipped with a photodiode array detector (PDA) for UV absorption was used. 100 µL of the irradiated toluene solution was injected into a Wakosil 5C18AR polymeric octadecyl silica (ODS) column and eluted by acetonitrile at a flow rate of 4 mL/min, which is transparent in the deep UV region down to 200 nm and thus advantageous for the observation of absorption features for separated species [14].

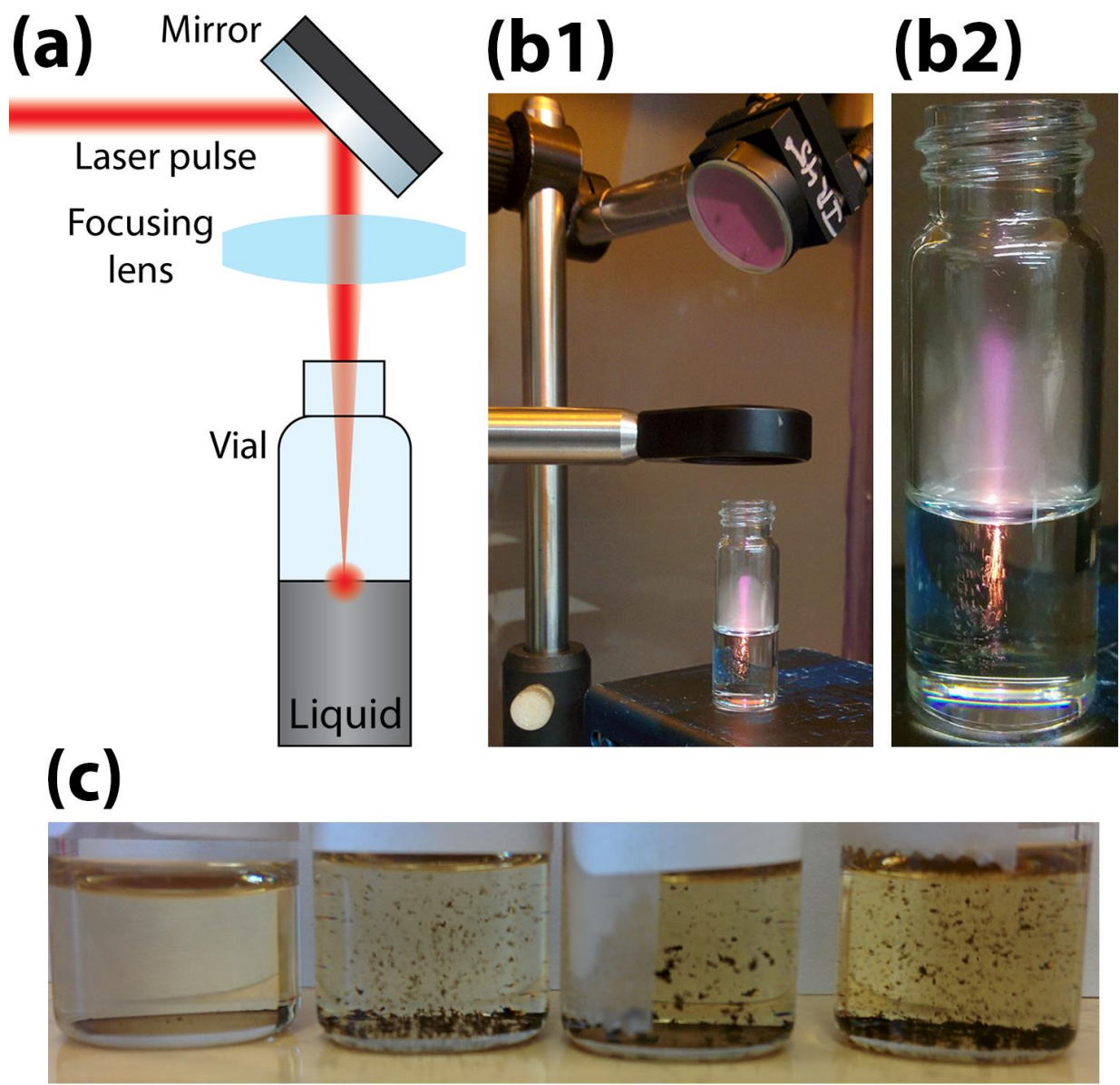

**Figure 1. (a)** schematic of the toluene irradiation experimental setup. **(b1)** a photograph of the experimental setup with the mirror, lens, and vial visible. The infrared laser pulses are not captured by the camera. **(b2)** a close-up of the toluene in the vial as it is irradiated. As the laser pulse enters the liquid it causes violent turbulence and generates a supercontinuum spectrum. Also, a small amount of liquid is ejected creating a region of toluene vapor that scatters the laser light, making it more visible to the camera. **(c)** samples irradiated for 1, 2, 3, and 4 hours respectively, starting from the leftmost vial. The volume of liquid in the vial remains constant throughout the irradiation.

## 3. Results

*3.1. Raman spectroscopy*

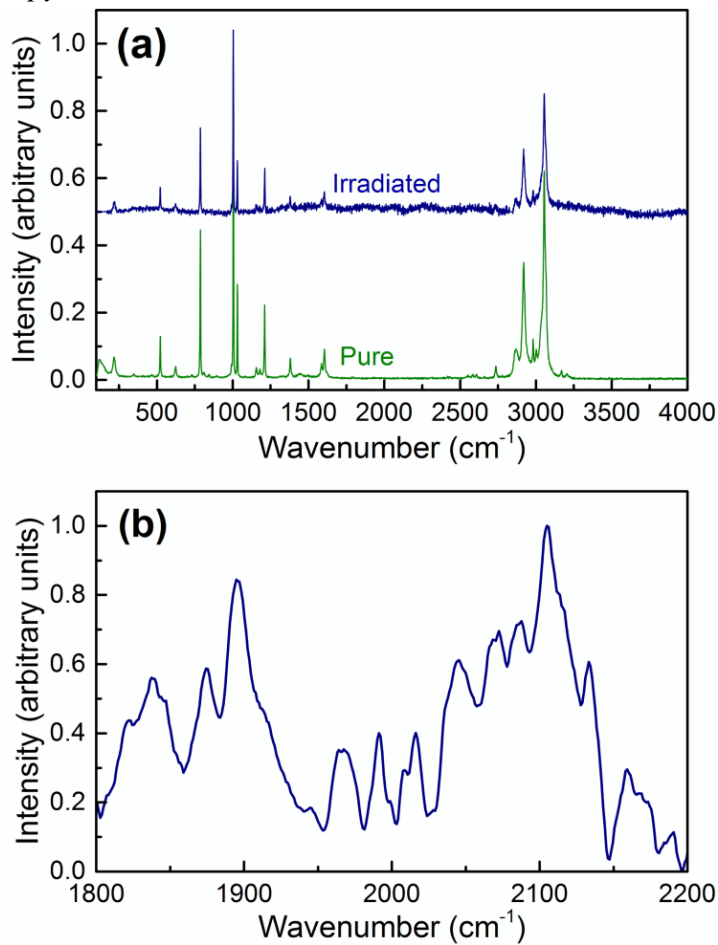

**Figure 2. (a)** The Raman spectra of pure (green) and irradiated (navy blue) toluene. The Raman spectrum of irradiated toluene originally contained a large background photoluminescence which was subtracted prior to plotting. Some of this background remains, resulting in small variations of the baseline. **(b)** A multi-accumulation scan of the 1800-2200 cm$^{-1}$ region in the spectrum of irradiated toluene showing over a dozen well resolved features indicative of polyyne formation.

Raman spectroscopy was employed to test for the presence of polyynes in the irradiated samples. A large background photoluminescence was observed in the Raman spectrum and was removed prior to analysis. Fig. 2(a) shows that there is no appreciable change in the energies of spectral features associated with the vibrational modes of toluene; however, a number of low intensity features do appear in the 1800-2200 cm$^{-1}$ region. The appearance of these peaks without the need of surface enhancement suggests a high concentration of linearly hybridized carbon, which may be a result of the extended irradiation time. A multi-accumulation scan of this area is presented in the inset of Fig. 2(b). Over a dozen well resolved features are shown, indicating a variety of polyyne and cumulene molecules have formed in the irradiated liquid [15,16]. Gaussian or Lorentzian deconvolutions were performed to determine the widths (full width at half maximum) and peak centers of the features. Not surprisingly, the strongest feature in this spectrum is centered at 2105 cm$^{-1}$. This mode generally occurs in acetylenic compounds with C≡C vibrations [17]. A number of chains were identified, the most significant of these being $C_{20}H_2$. Density functional simulations suggest that this molecule has four totally symmetric vibrational modes in the polyyne fingerprint region [16]. These match to within a wavenumber with four features found in the present sample at frequencies of 2188, 2133, 2086, 1911 cm$^{-1}$. To date this would be the longest polyyne molecule fabricated using the femtosecond liquid interaction method for polyyne synthesis. The presence of spectral features associated with $C_{18}H_2$, $C_{16}H_2$, $C_{14}H_2$, $C_{12}H_2$ and $C_{10}H_2$ have also been identified in these samples [15,18,19].

In samples with many spectral features such as this one, the identification of specific molecular species can be quite challenging due in part to the fact that there is often overlap of the fundamental vibrational modes. For example, it is well established that as the length of the polyyne chain increases, the frequency of its primary vibrational mode decreases [15]. This may result in the primary modes of long polyynes overlapping modes that belong to shorter cumulenes. To conclusively confirm the presence of these polyynes, high-performance liquid chromatography was performed.

*3.2. High-performance liquid chromatography*

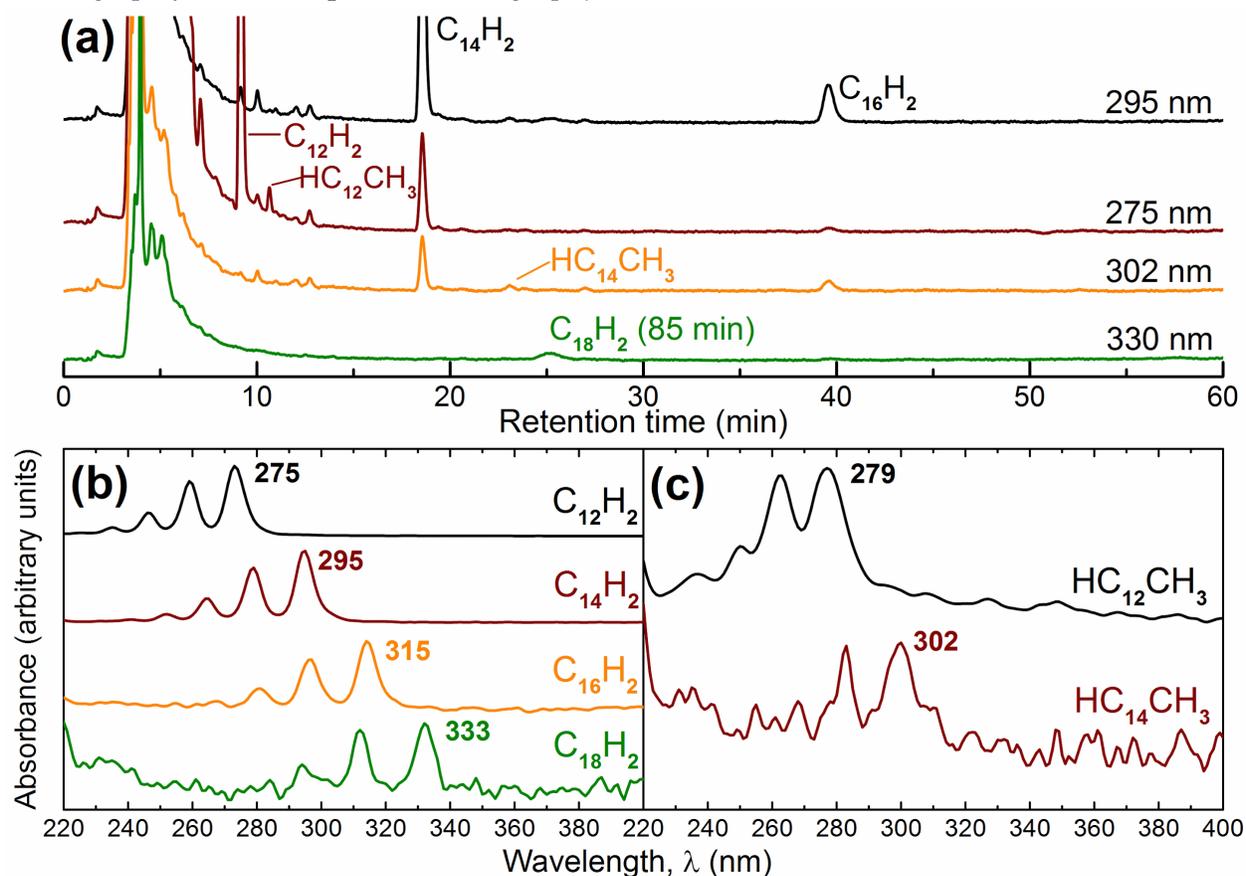

**Figure 3. (a)** HPLC chromatograms taken at various absorption wavelengths. Peaks corresponding to each polyyne are labeled. The $C_{18}H_2$ peak shows up one analytical cycle later at 85 mins. The wavelengths were chosen to showcase all possible synthesis products using the minimum number of curves. **(b)** UV-Vis absorption spectra of individual hydrogen-capped polyynes, $H(C\equiv C)_nH$ (*n* = 6-9), from separated fractions extracted through HPLC with the maximum absorption peaks labeled in nanometers. Each absorption spectrum is normalized such that the baseline (zero absorbance) is at zero while the maximum value is 1. The spectra are offset for clarity. **(c)** Same as (b) but for methyl-capped polyyne fractions, $H(C\equiv C)_nCH_3$ (*n* = 6,7). These spectra were extracted for a sample irradiated for 3 hours.

We employed HPLC to distinguish the signal from different molecules in our irradiated samples. During the separation, multiple UV absorption spectra were recorded from which the chromatogram in Fig. 3(a) was produced. The peaks corresponding to each polyyne are labeled. Note that the $C_{18}H_2$ peak shows up one analytical cycle later at 85 mins. The wavelengths were chosen to showcase all possible synthesis products using the minimum number of curves. The absorption spectra in Fig. 3(b) was also produced from the HPLC data, confirming the presence of hydrogen-capped polyynes from $C_{12}H_2$ up to $C_{18}H_2$. The observed absorption features and their peak maxima in nanometers is shown in Fig. 3(b) for hydrogen-capped polyynes, $H(C\equiv C)_nH$ ($n$ = 6-9). The maximum absorption peaks are labeled in nanometers. Each absorption spectrum is normalized such that the baseline (corresponding to zero absorbance) is at zero while the maximum value is 1 and the spectra are offset for clarity. Fig. 3(c) shows the observed absorption features and their peak maxima in nanometers for methyl-capped polyynes, $H(C\equiv C)_nCH_3$ ($n$ = 6,7). They are identical with previous observations [9,20]. The spectra shown in Fig. 3 were extracted from a sample irradiated for 3 hours. The redshift in peak absorption wavelength empirically follows a decaying exponential relationship with increasing polyyne length and eventually saturates at 485 nm for $n \approx 48$ [1]. Redshifts in the absorption spectra of hydrocarbon molecules generally occur due to increasing bond conjugation which in turn results in a greater number of delocalized π-electrons present within the system [21]. This results in less energy being required for an electronic transition, and therefore an absorption peak is observed at a longer wavelength. Since the only appreciable change in the Raman spectrum of irradiated toluene is the appearance of polyyne modes it is reasonable to assume that the shift in the absorption spectrum is related to the formation of these substances. In fact, a relationship in which the HOMO-LUMO energy gap decrease with increasing chain length has been established for polyynes synthesized using other techniques [22,23].

The toluene and shorter polyynes are responsible for the peak between 4-6 minutes in Fig. 3(a) especially with the high absorbance of toluene throughout the UV region. The shorter polyynes exhibit a higher signal-to-noise ratio in their absorption spectra in Figs. 3(b) and 3(c) indicating a decrease in polyyne concentration and rate of synthesis as the length of the polyyne increases. This suggests a synthesis description based on spontaneous sequential addition reactions between the dissociation products of toluene. Of note is that the signal-to-noise ratio between the methyl-capped polyynes and $C_{18}H_2$ is similar suggesting that methyl-capped polyyne concentrations need not be much lower than those of hydrogen-capped polyynes despite the added complexity of a methyl cap. An absorption peak belonging to $C_{20}H_2$ was not observed, possibly due to it being indistinguishable from noise given the already low signal-to-noise ratio on the $C_{18}H_2$ absorption spectrum.

It is very likely that shorter polyynes exist in high concentrations as previous works have synthesized large amounts of polyynes as short as $C_6H_2$ in hexane for example [12]. However, in this work their peaks would appear earlier than $C_{12}H_2$ (~9 minutes) in the HPLC chromatogram, Fig. 3(a), and thus coincide with the highly prominent toluene peak (3-7 minutes), making it difficult to identify any shorter polyynes (or any non-toluene peak in that region) with the current setup. It is also very possible that methylpolyynes shorter than $C_{12}CH_3$ are present with higher concentrations but their chromatographic peak would be quite close to the toluene peak (3-7 minutes), again making it very difficult to identify any.

## 4. Discussion

Polyyne production by the direct ultrafast laser irradiation of acetone, n-hexane, n-decane, and benzene has been studied in prior work. Hu et al. [11] described the irradiation of 20 mL of acetone using 90 fs, 300 μJ, 800 nm, 1 kHz laser pulses focused by a 40 mm focal length lens for an unreported amount of time. Polyynes as short as $C_6H_2$ were observed using surface enhanced Raman spectroscopy. Without HPLC longer polyynes could not be conclusively identified, although the presence of $C_8H_2$-$C_{14}H_2$ was suspected. Sato et al. [12] irradiated ~5 mL of n-hexane and n-decane using 100 fs, 0.9 mJ, 800 nm, 1 kHz laser pulses focused by a 50 mm focal length lens for 30-300 minutes, and produced hydrogen-capped polyynes up to $C_{12}H_2$. It was reported that irradiation of both n-hexane and n-decane produced the same types of polyynes, with a slightly higher concentration of polyynes produced during the irradiation of n-hexane. Thus it was concluded that single step hydrogen elimination from the parent hydrocarbon was an unlikely mechanism for polyyne synthesis. Wesolowski et al. [13] irradiated 3.5 mL of benzene using 120 fs, 300 μJ, 800 nm, 1 kHz laser pulses focused by a 40 mm focal length lens for 1.5 hours. Both amorphous carbon nanoparticles and hydrogen-capped polyynes up to $C_{14}H_2$ were produced and a multi step carbon addition process was suggested as the polyyne synthesis mechanism. It is important to note that in these previous experiments, the laser was focused deep below the surface of the liquid solvent, deep enough to contain the laser filament produced through self-focusing and reported to be 2-3 cm in length by Sato et al. [12]. Herein it was found that focusing the laser at the meniscus of the liquid solvent produced a much greater concentration of polyynes and in particular, the longest polyynes and the methyl-capped polyynes were not observed until we focused at the surface.

Perhaps not surprisingly, in previous work, only hydrogen capped polyynes have been observed as a product of ultrafast laser irradiation of liquid solvents. This is likely due to the fact that the previously targeted liquid solvents did not contain functional groups. Toluene consists of a methyl group attached to a phenyl group and is a mono-substituted benzene derivative The ionization potential of toluene (8.82 eV) is lower than that of benzene (9.24 eV), and it has been shown to readily fragment in the gas phase [24,25]. In the liquid phase the clamping intensity of a laser filament in toluene is estimated to be nearly twice that of benzene ($I_T/I_B \approx 1.8$) suggesting that there should be enough energy inside the focal regions of the filament to initiate dissociation [26–28]. The dissociation of toluene produces many fragments including, $CH_n^+$ $C_3H_n^+$, $C_4H_n^+$, and the singly and doubly charged parent ions $C_7H_8^+$ and $C_7H_8^{2+}$ [24,25]. Importantly, many of these ions still contain the methyl group. It has been shown here that both hydrogen capped polyynes up to $C_{18}H_2$ and methyl capped polyynes up to $HC_{14}CH_3$ are produced as a result of the ultrafast laser irradiation of toluene. The formation of polyynes and methylpolyynes that contain a larger number of carbon atoms than the parent molecule implies that dissociation of the parent must be taking place and supports the previously suggested carbon addition process as a mechanism for polyyne formation [29].

The irradiation process may also produce residual compounds such as benzene, acetylene, or butadiyne. Gas chromatography-mass spectroscopy (GC-MS) may be used to detect these compounds in the irradiated samples due to its high detection sensitivity in mass spectrometry. An example using $HC_9N$ has been reported on [14] and methylpolyynes have been identified by GC-MS in previous works [20]. Future investigations of how the laser pulses' properties affects polyyne production rates during the irradiation process could lead to significantly greater polyyne yields and should provide further opportunities for characterizing these residual compounds, for example using $^{13}$C-NMR and H-NMR once milligrams of the synthesis products can be isolated.

It is generally accepted that the stability of polyynes is greatly improved by the addition of end-caps [30,31]. These groups prevent the reactive polyyne cores from cross-linking or reacting with other species by providing steric bulk and electronic stabilization [31]. The end-cap control demonstrated herein significantly improves the applicability of the ultrafast laser liquid irradiation polyyne synthesis technique and opens up the possibility of synthesizing more complex polyyne molecules. While the generation of polyynes is our main focus, our method is also very versatile and could be applied to many liquid chemical systems. While attempting to synthesize $CH_2$-capped carbon chains is tempting, to test the hypothesis that such polyynes form a magnetic semiconductor as theorized by Liu et al. [3], $CH_2$ capped carbon chains must be cumulenic (=C=C=$CH_2$) which have small band gaps and thus are easily excited by photons and heat at room temperature. If any such cumulenes are produced in the irradiation process, they would decay quickly as transient species well before any HPLC analysis may be carried out. However, characterizing the possible capped polyynes that may be produced using this technique will be interesting as end-cap modification is difficult in practice and could lead to the production of polyynes with desirable electrical and optical properties.

## 5. Conclusion

Polyynes were produced by femtosecond laser pulse irradiation, directly initiating synthesis from a liquid containing organic molecules without the need to introduce any carbon particles. Raman spectroscopy of irradiated samples suggested the presence of polyynes up to $C_{20}H_2$ and $HC_{14}CH_3$. From HPLC results we were able to confirm the presence of hydrogen-capped polyynes up to $C_{18}H_2$ and additionally identify methyl-capped polyynes up to $HC_{14}CH_3$. The mechanism of polyyne formation from the parent molecule is unclear but single step rearrangement of the parent molecule is not likely, and impossible for polyynes longer than $C_6H_2$ for the case of toluene. The presence of methylated polyynes indicates the importance of incomplete fragmentation at least in end-cap formation. Our result indicates a route to the control of end-cap populations by selection of the initial solvent or mixture of solvents to be irradiated.

## Acknowledgements

We gratefully acknowledge research funding from NSERC (Natural Sciences and Engineering Research Council of Canada). A.R. was supported by the Research Internship Program of Tokyo Metropolitan University. T. W. was supported by the MEXT-Supported Program for the Strategic Research Foundation at Private Universities. We would also like to thank the anonymous reviewers for their helpful suggestions.